# High Fidelity Human Trajectory Tracking Based on Surveillance Camera Data

Ze-Xu Li[1] and Lei Fang [*1]

[1]Department of Civil and Environmental Engineering, University of Pittsburgh, Pittsburgh, Pennsylvania 15261, USA

December 26, 2023

**Abstract**

Human crowds exhibit a wide range of interesting patterns, and measuring them is of great interest in areas ranging from psychology and social science to civil engineering. While *in situ* measurements of human crowd patterns require large amounts of time and labor to obtain, human crowd experiments may result in statistics different from those that would emerge with a naturally emerging crowd. Here we present a simple, broadly applicable, highly accurate human crowd tracking technique to extract high-fidelity kinematic information from widely available surveillance camera videos. With the proposed technique, researchers can access scientific crowd data on a scale that is orders of magnitude larger than before. In addition to being able to measure an individual's time-resolved position and velocity, our technique also offers high validity time-resolved acceleration and step frequency, and step length. We demonstrate the applicability of our technique by applying it to surveillance camera videos in Tokyo Shinjuku streamed on YouTube and exploiting its high fidelity to expose the hidden contribution of walking speed variance at the crossroad. The high fidelity and simplicity of this powerful technique open up the way to utilize the large volume of existing surveillance camera data around the world for scientific studies.

Crowd Dynamics, Human Trajectory Tracking

## 1 Introduction

**Q**uantitative measurement of naturally emerging crowds is of great importance in fields ranging from cognitive science [1, 2] and social science [3] to civil engineering [4, 5, 6, 7]. Human trajectories can yield rich information, for example, on understanding individuals' perception and cognition [1, 2, 8, 9] as well as their collective behaviors [10, 11, 7, 12], contributing to the development of strategies for improving transportation efficiency and safety [13, 14, 15, 6, 16]. Detailed crowd information is revealed from the trajectories of individuals as well as the time resolved velocity and acceleration and step frequency and step length along the trajectories.

Trajectories in crowds are generally measured by video cameras and undertaken in two major ways: lab observations [17, 18, 19, 16] and *in situ* measurements [6, 20]. However, while lab observations with volunteers offer high fidelity measurement and well-controlled crowds, the crowds in the lab may behave differently from the ones that emerge naturally [21]. Furthermore, lab measurements are usually conducted in a small area with a limited number of individuals, making measurements easier to obtain. *In situ* measurements, on the other hand, require painstaking on-site calibration and costly field trips. In addition, both lab observation and *in situ* measurements can only generate a limited volume of crowd trajectory data. These limitations of existing measurement techniques impede at least three major areas of crowd research due to the limited amount of crowd kinematic data. First, transportation safety and comfort studies in traffic systems require a large volume of *in situ* human trajectory data over a wide time span [22, 23]. Second, the study of coordinated acts of intentional physical violence in human crowds [24, 25] suffers from a limited amount of real quantitative data. However, many coordinated acts of intentional physical violence are accidentally captured by the surveillance cameras, and our technique can extract quantitative kinematic information from surveillance camera data. Third, it is an urgent need to understand crowds with social distances for better crowd management for a potential pandemic [13, 17]. An individual's behaviour in socially distanced crowds in different kinds of built environments is not well-understood to date due to the limited measurement techniques.

[*]e-mail address: lei.fang@pitt.edu

Our technique offers a high fidelity and cost-effective framework (see Fig. 1) that utilizes surveillance camera data, which is taken and archived every second and is widely available all over the world [26]. Our technique opens up the new possibility of extracting high fidelity kinematic information from surveillance camera data, making available a novel way to investigate human behavior patterns in crowds. There are three major difficulties in utilizing the surveillance camera data for human trajectory tracking. First, human identifications with a non-uniform background are challenging. In typical experiments and *in situ* measurement, researchers measure crowds with relatively uniform backgrounds and specifically choose to measure crowds in which individuals have similar clothing [6, 20]. For example, in volunteer experiments, participants are usually required to follow a specific dress code to reduce the tracking difficulty, such as wearing red hats with black suits for ease of head identification [17]. Second, most surveillance cameras are not calibrated, and the parameters of the cameras, both intrinsic and extrinsic, are unknown. For example, contrary to a typical camera setup in a lab and *in situ* where the camera's line of sight is usually perpendicular to the ground [6, 17], surveillance cameras' line of sight is usually at an unknown angle to the ground. As a consequence, a researcher who wants to utilize an archived surveillance camera video cannot get the positions of the individuals in the physical space in the video recording. Third, the image domain of a surveillance camera usually covers a big area, and the spatial resolution of the resulting video is relatively coarse. For instance, typically, one pixel in a surveillance video can correspond to around a centimeter in the physical space, which is one order of magnitude coarser than resolutions in lab experiments.

Our novel technique overcomes these three difficulties and offers a simple yet powerful comprehensive framework (Fig. 1) to extract high fidelity kinematic information (position, velocity, acceleration, step frequency, and step length) from surveillance cameras. First, the human identification issue is overcome by utilizing a mask region-based convolutional neural network (R-CNN) [27]. Mask R-CNN allows the accurate determination of humans even in videos with complex backgrounds. Second, the calibration issue is resolved using a modified direct linear transformation technique (DLT) [28]. With DLT, we only need to physically visit the area of interest once and measure the relative positions of four fixed points. These four points are sufficient for calibrating all the camera videos in that area. Third, despite the relatively coarse spatial resolution, we conducted a comprehensive error analysis for the resolution issue and found that our measurement framework can reliably extract high fidelity kinematic information of an individual's trajectory even at relatively compromised spatial resolutions.

An example of the value of our tracking framework is demonstrated when, using video data from a remote surveillance camera at Tokyo Shinjuku with unknown camera parameters and location, we are able to extract high validity kinematic information and use the time-resolved kinematic information along a trajectory to reveal that an individual's speed change is largely correlated to step length rather than step frequency. This result reveals that transportation facilities, such as escalators and revolving gates, that may trigger step frequency change need to be replaced by alternative designs that trigger step length change.

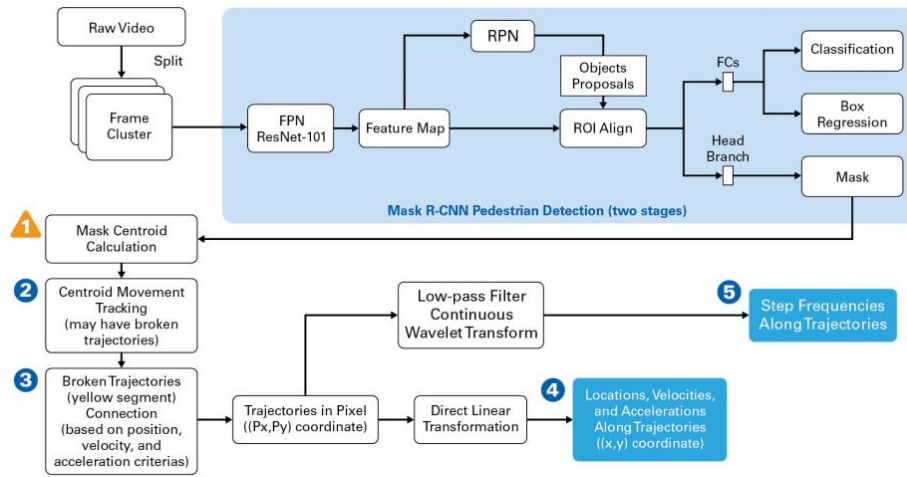

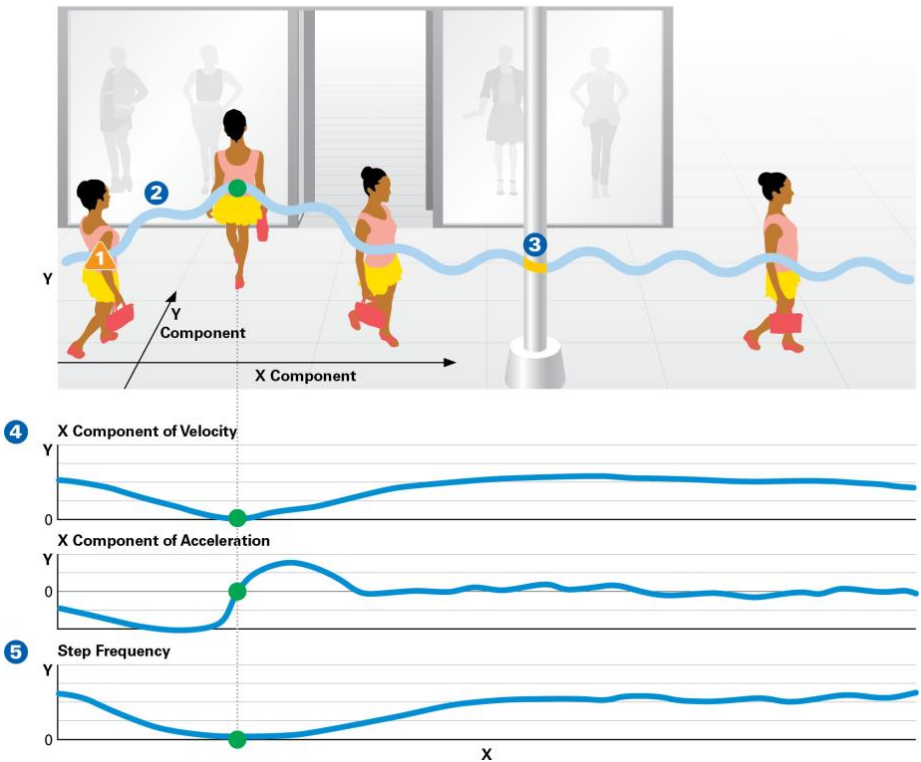

Figure 1: Top: a flow chart of our novel human crowd tracking technique. The raw video is fed into a mask R-CNN, which outputs the mask of the human in the image. Then the centroids of the masked human are calculated and tracked. Any incidental broken trajectories are connected. Using low-pass filtering and continuous wavelet transformation, the technique gets time-resolved step frequencies. The time-resolved position, velocity, acceleration and step length can be obtained via direct linear transformation as well as Gaussian derivative kernels. Middle: a schematic of measuring a woman doing window shopping. She walked toward the window and stopped momentarily. Then, she continued to walk, and she was partially blocked by a utility pole. Bottom: the schematic of X component of velocity, acceleration, and step frequency.

This result can inform transportation system design to enhance the safety and comfort of pedestrians.

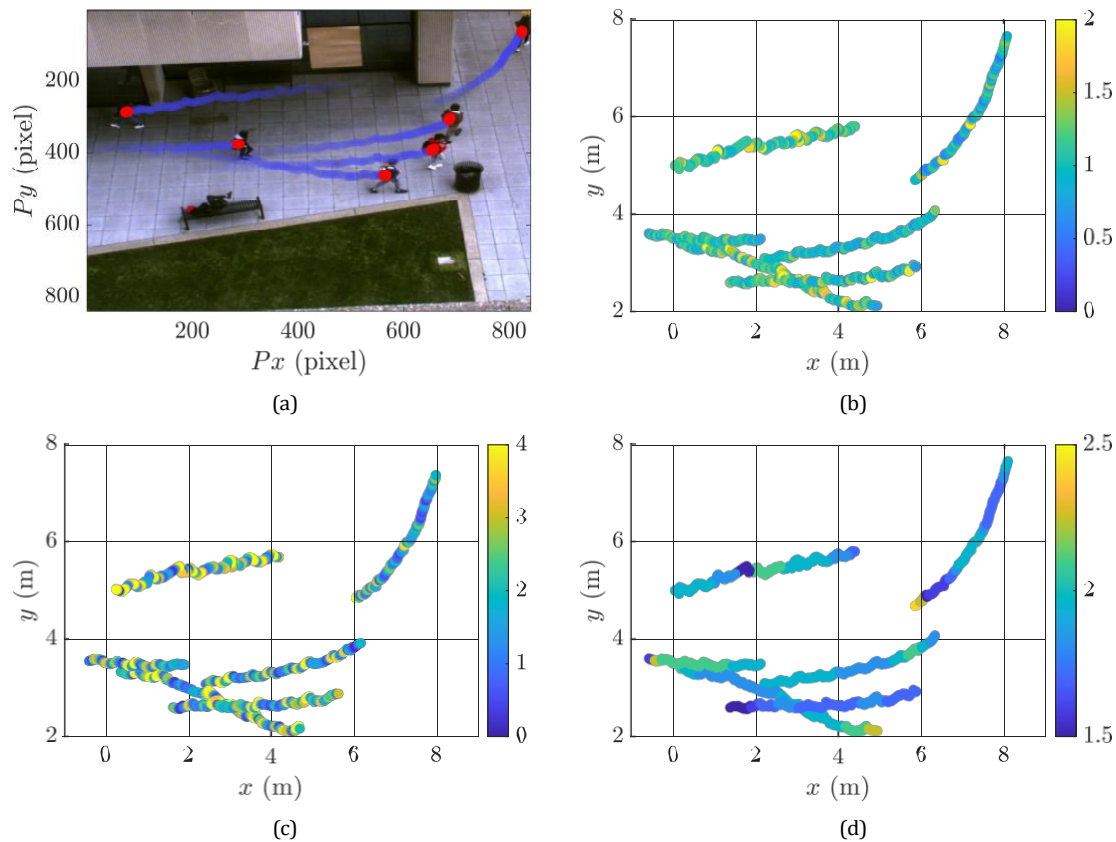

Figure 2: **(a)** A snapshot of individuals with measured trajectories in pixel coordinate in the past 6 seconds. The centroids are indicated as red dots. **(b)** Corresponding trajectories in the physical space (x-y coordinate) after direct linear transformation. The trajectories are colored by velocity. **(c)** Same trajectories as the ones in (b) colored by acceleration. **(d)** Same trajectories as the ones in (b) colored by step frequency.

## 2 Methods

### 2.1 Tracking in pixel coordinate

To handle the issue of individual identification in a non-uniform background, we used the method called mask region-based convolutional neural networks (Mask R-CNN). Mask R-CNN is one of the highest-performance single-model techniques in image segmentation and instance segmentation. Developed on top of Faster R-CNN, mask R-CNN not only outputs for each candidate object a class label and a bounding-box offset but also an object mask [29, 30]. Here, we used Mask R-CNN as our first step to successfully mask out the boundaries of humans in the image [27].

After getting the mask of the human, we calculated the centroids of the mask for each human in each frame (Fig. 1, label 1) and identified the centroids as the point position of the humans. During the reconnection of the broken trajectories, the correspondence of the centroids between frames was determined using the Lagrangian particle tracking technique [31, 32]. The Lagrangian particle tracking resulted in the trajectories of individual centroids in the camera's pixel coordinate (Fig. 1 label 2; Fig. 2a). We call the horizontal axis $\mathbf{p_x}$ and vertical axis $\mathbf{p_y}$.

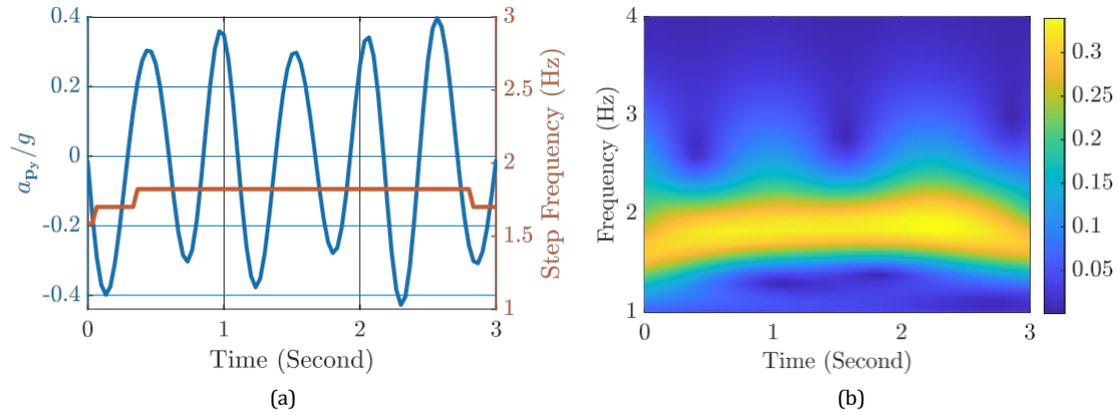

Figure 3: **(a)** Vertical acceleration ($a_{p_y}$) as a function of time with measurement from a camera with a 25 mm lens. **(b)** Power plot obtained from continuous wavelet transform of vertical acceleration ($a_{p_y}$), where maximum power at each time corresponds to step frequency.

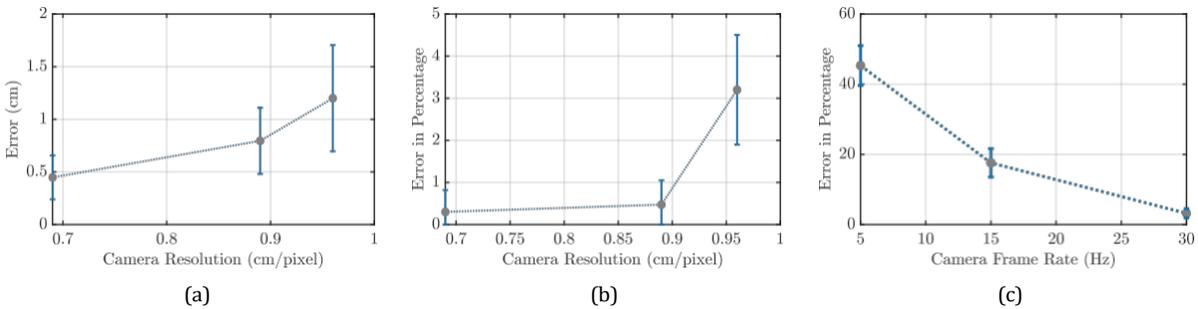

Figure 4: **(a)** The errors in direct linear transformation for cameras with different resolutions. **(b)** The errors in step frequency normalized using the mean ground truth step frequency for cameras with different resolutions. The step frequency is averaged over every 0.2 second interval. **(c)** The error in step frequency normalized using the mean ground truth step frequency for cameras with different image rates. The error bars represent 1 standard deviation.

Despite the high accuracy of mask R-CNN [29], it will inevitably miss some of the humans in certain frames, which will lead to broken trajectories. In addition, some objects, such as utility poles, can also lead to broken trajectories (Fig. 1, label 3). In order to connect a broken trajectory that corresponds to the same person, we developed a set of criteria:

1) Any two trajectories that overlap in time are not connected because one person cannot present at two positions at the same time.

2) Any two trajectories that have a time gap (from the end of one prior trajectory to the beginning of the following trajectory) longer than a human's relaxation time (0.5 s) [33] are not connected. Relaxation time is the time it takes for a human to make a nonlinear change to one's trajectory. If the missing information is longer than a human's relaxation time, the missing information will not be recoverable due to an unknown nonlinear change in the trajectory.

3) If an individual's speed is greater than 90 percentile of the population's speed during the potential connection, the trajectories are not connected. The excessive speed due to the re-connection suggests that the connection is not physical.

Since the mask R-CNN has extremely high accuracy in detecting human positions, most broken trajectories have short gaps between them, and our criteria can efficiently connect the trajectories. We used linear interpolation for the missing positions because the broken time is shorter than the human's relaxation time, and any nonlinear change of trajectory would then be negligible.

## 2.2 Calculation of step frequency

As we observe in Fig. 2a, the trajectories in the pixel coordinate are sine-wave-like, and these sine-wave-like trajectories can help us to determine the time-resolved step frequencies (Fig. 2a). We used the accelerations in a vertical direction in pixel coordinate ($a_{\mathbf{p_y}}$) to extract step frequencies. In Fig. 3a, we see $a_{\mathbf{p_y}}$ vary along a trajectory is sine-wave-like which corresponds to an individual's step frequencies. The time variation in step frequencies was calculated by applying a continuous wavelet transform (CWT) [34] (Fig. 3b). For each time, the power of different frequencies is shown, and the maximum in power corresponds to the step frequency (Fig. 1, label 5) [35]. The acceleration magnitude of $a_{\mathbf{p_y}}$ is not necessarily the acceleration magnitude in the vertical direction in physical space. However, here we only need the acceleration variation frequency in $\mathbf{p_y}$ direction for calculating the step frequency, and the magnitude is not important.

## 2.3 Calculation of position, velocity, acceleration, and step length

To this point in the examination of data, the trajectories were located in pixel coordinates (Fig. 2a). However, in order to scientifically understand the crowd, we used DLT to transform the trajectories from pixel coordinates to physical space (Fig. 1, label 4; Fig. 2b). Assuming the area of interest has flat ground, the DLT only requires one camera and four calibration points (Supplementary Material; Sec. Two-dimensional Direct Linear Transformation). The assumption of flat ground is valid in most city areas, which is especially true over a relatively small region of interest.

After the calibration, we can get the trajectories in physical space and, by taking derivatives along trajectories, we can calculate the velocities and accelerations along the trajectories. However, the steps of individuals' walking will lead to undesired high-frequency perturbations along their trajectories after DLT. To characterize the motility of people, velocity ($\mathbf{v}(t)$; Fig. 2b) and acceleration ($\mathbf{a}(t)$; Fig. 2c) along the trajectories were derived from the position information through Gaussian derivative kernels [36]. Gaussian derivative kernels were widely used in experimental research of Lagrangian turbulence to filter out high-frequency noise [37, 38, 39] (see Methods). After applying Gaussian derivative kernels, smoothed velocity and acceleration along the trajectories can be calculated. The velocity and acceleration can reflect the mobility of the human body while the step-induced oscillations are filtered. Moreover, the time-resolved step length can also be calculated based on the fact that instantaneous velocity is the multiplication of step length and step frequency.

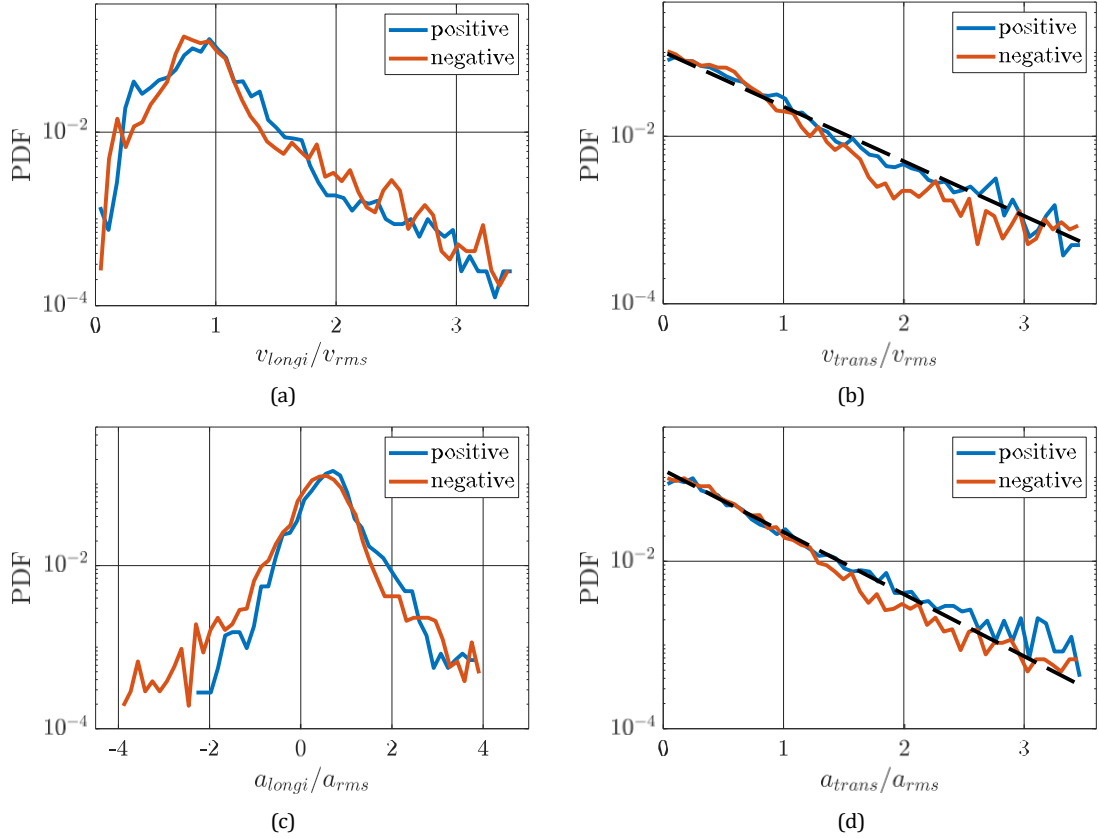

Figure 5: **(a)** PDF of normalized longitudinal velocity; the root mean square (RMS) velocities used for normalization of positive and negative direction are 2.03 m/s and 2.13 m/s, respectively. Positive and negative refers to the individuals walking from two opposite sides of the zebra crossing. **(b)** PDF of normalized transverse velocity, the RMS velocities used for normalization of positive and negative direction are 0.23 m/s and 0.29 m/s, respectively. **(c)** PDF of normalized longitudinal acceleration; the RMS accelerations used for normalization of positive and negative direction are 1.28 m/s² and 1.25 m/s², respectively. **(d)** PDF of normalized transverse acceleration; the RMS accelerations used for normalization of positive and negative direction are 0.97 m/s² and 0.98 m/s², respectively.

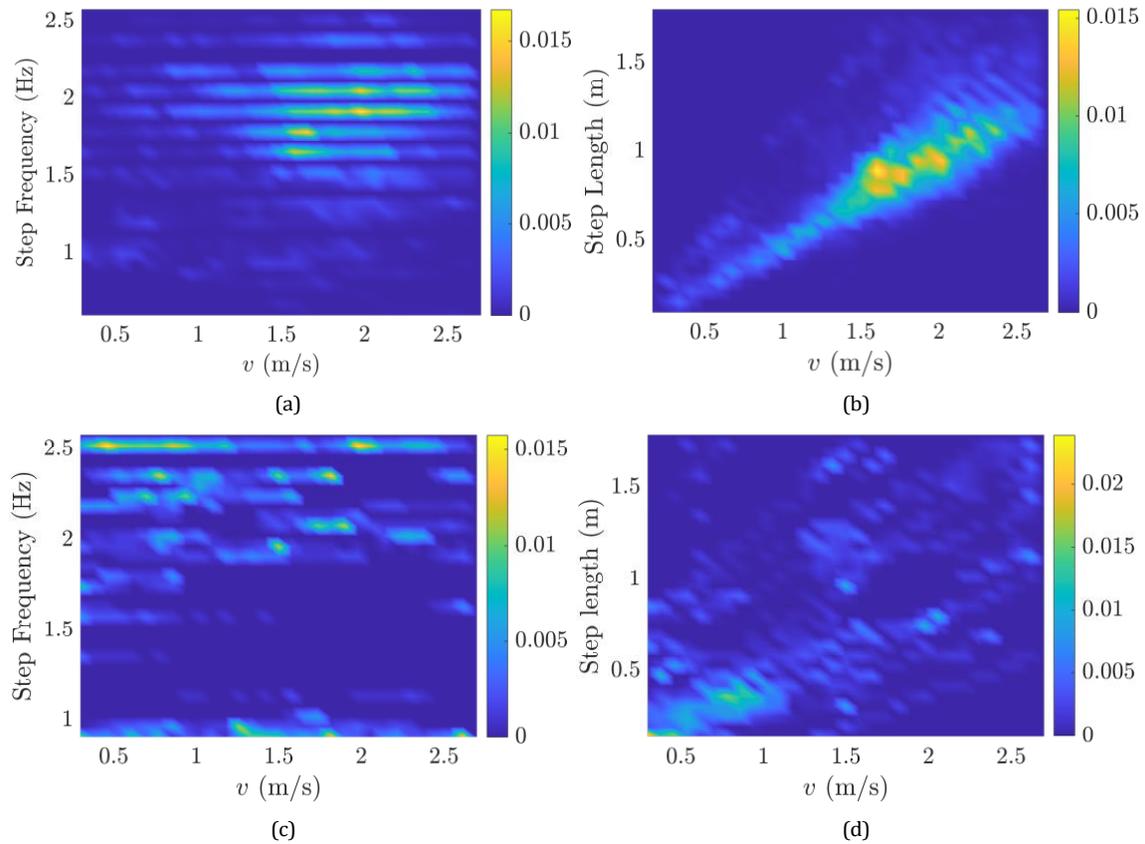

Figure 6: **(a)** A joint PDF of velocity magnitude and step frequency. **(b)** A joint PDF of velocity magnitude and step length. **(c)** A joint PDF of velocity magnitude and step frequency averaged over 10 walkers. **(d)** A joint PDF of velocity magnitude and step length averaged over 10 walkers (same as (c)).

Up to now, we have described all the necessary steps to extract time-resolved position, velocity, acceleration, step frequency, and step length from a surveillance camera. This technique is extremely useful for gaining valuable data for a variety of scientific areas, such as social science, psychology, physics, and civil engineering because this technique enables us to extract scientifically sound crowd trajectory data from the enormous amount of video data archived world wide. We only need to find the relative distance between four calibration points to conduct calibration. For instance, researchers who are interested in coordinated acts of intentional physical violence in crowds can extract high validity kinematic information from the target video using four calibration points at the target site. With our technique, the researchers can access scientific crowd data on a scale that is orders of magnitude larger than before.

## 2.4  Error Analysis

We now examine the level of error that may be encountered using our novel technique. First, we will quantify the uncertainty of DLT under different resolutions. Second, we will determine the step frequency accuracy by comparing the step frequency from our novel technique and the step frequency from an accelerometer. Since step frequency is derived from the second-order derivative of the trajectories, if the step frequency is accurate, the lower order statistics, such as position and velocity, must be accurate.

We used a Blackfly color camera with Fujinon 12.5, 16, and 25 mm lenses, which correspond to spatial resolutions of 0.96, 0.81, and 0.69 cm per pixel, respectively. The relatively coarse resolutions mimic the coarse resolution of a surveillance camera. We imaged at the $7^{th}$ floor of Benedum Hall on the University of Pittsburgh campus, and the image region is at the ground level with square bricks of 59.0 *cm*× 59.0 *cm* (Fig. 2) as background.

For quantifying the uncertainty of DLT, we hand-picked a grid of brick corners to the best of our ability and transformed the points from pixel coordinates to physical coordinates using DLT. With the known brick size, we know that the transformed points should be on a square grid, and each point should be 59.0 cm away from its neighbors (ground truth). By comparing the transformed points and the ground truth, we determine the mean error for different spatial resolutions, which is shown in Fig 4a. The uncertainty of a hand-picked point should be on the order of 1-pixel [31]. We see that the transformed error is, in fact, on the same order of magnitude as camera spatial resolution, which shows that our DLT method is of high accuracy and that the main source of spatial uncertainty comes from the finite camera resolution.

To determine the uncertainty of step frequencies, we simultaneously measured a volunteer's step frequency using both a camera and an accelerometer. The accelerometer recorded the three-component acceleration at 30 Hz, and we used the CWT to process the acceleration in the direction of gravity to get the ground truth step frequency. The uncertainty of the step frequency derived from the video image is evaluated for different spatial and temporal resolutions of the camera. We varied the spatial resolution using the three lenses mentioned above and varied the temporal resolution using different frame rates: 30, 15, and 5 frames per second.

With the simultaneous measurement of step frequency by a camera and an accelerometer, we first calculate the mean step frequency every 0.2 seconds and calculate the difference between the two time series. The error in step frequency is normalized by the mean ground truth step frequency. We find that our technique is robust for cameras at different resolutions. Given a spatial resolution of 0.69 cm per pixel, the mean error in step frequency is within 0.5%. Even at around 1 cm per pixel spatial resolution, the mean error in step frequency is only around 3%. One needs to note that, in general, step frequency is around 2 Hz, and we are trying to resolve and validate the step frequency at 5 Hz. If we compare the performance between video recording and human counting for step frequency analysis, step frequency from video recording is more accurate with a much larger temporal resolution.

We also find that step frequency measurement is more sensitive to the camera's temporal resolution than to spatial resolution. Due to the Nyquist limit, the camera's image acquisition rate should be at least twice the step frequency [40]. This means that the step frequency measurement should be good as long as the camera acquires images at 4 frames per second. However, we find the measurement error was extremely large at 5 frames per second. This may be caused by the interplay between relatively low temporal and spatial resolution. For most modern cameras, the frame rate can easily reach 30 frames per second, meaning the step frequency measurement should be good for most modern cameras.

## 2.5 Lagrangian statistics for the crowd at a crossroad

We claim that our technique can be extremely useful for several areas of study: (1) walking safety and comfort study in traffic systems, (2) coordinated acts of intentional physical violence in human crowds, and (3) behavior of socially distanced crowds in a large gathering. To demonstrate the usability of our technique, we used our technique to extra Lagrangian statistics for crowds at a crossroad in Tokyo Shinjuku. We specifically studied individuals' kinematics at a zebra crossing. First, we calculated the mean velocity of individuals at the zebra crossing. The mean velocity is roughly perpendicular to the zebra line. Next, we projected the instantaneous velocity/acceleration to the directions parallel and perpendicular to the mean velocity direction, calling the projected velocity/acceleration longitudinal and transverse velocities/accelerations, respectively.

For the probability density function (PDF) of longitudinal velocity $v_{longi}$ (Fig. 5 a), results show a maximum probable value of around 0.8 in the two opposite directions, which means people like to cross the road with a velocity of 1.6 m/s approximately. In contrast, the longitudinal acceleration $a_{longi}$ (Fig. 5 c) distribute on the two sides of the zero value with a maximum probability value at around 0.65 and a mean acceleration magnitude 0.84 m/s$^2$. The acceleration PDF shows that there are both accelerating and decelerating individuals, but pedestrians tend to have more acceleration in their walking direction. The deceleration is due to the maneuvering when two opposite groups encounter each other. Moreover, we see that the transverse velocity and acceleration are distributed roughly exponentially. For our data, the mean and standard deviation of step frequency was 1.9 Hz and 0.37 Hz, respectively, consistent with previous studies [21, 41].

Using our novel technique, we extracted archived surveillance camera data across seasons, from September 2020 to August 2021, in Tokyo Shinjuku; we can report that the velocity acceleration, step frequency, and step length statistics are not statistically different for summer and winter (p-value < 0.05; Fig. S3), i.e., there is no seasonal variation of kinematics for pedestrians. However, our data reveals that for naturally emerging crowds on a zebra crossing, the variation in an individual's speed is highly correlated with step length but not correlated with step frequency (Fig. 6 a and b). A similar conclusion has been derived before[42, 43]. This indicates that the population's speed variance originates from the length of their steps rather than from how quickly they take steps.

Due to the limits of the measurement technique, what is not known to date is whether the variance in step length is among individuals or within the same individual and the variation is over time. With our simultaneous measurement of time resolved position, velocity, acceleration, step frequency and step length, we can reveal this mystery. The positive correlation between individuals' speed and step length has two possibilities. The first is that velocity variation is mainly among different individuals and that different step length is coupled with individuals' physiological properties, such as height. If this is true, the designer of the transportation system needs to take into account the distribution of the step length among populations because, naturally, pedestrians don't adjust their step length too often. The second possibility is that the speed can vary noticeably for any individual, and the individuals change their speed by means of changing step length. To answer this question, we calculate the joint PDF of an individual's speed and step length for one individual. Since the data point for one individual is limited, we use Bootstrapping [44]. Fig. 6 c and d show the averaged joint PDF for one individual, and the average is taken over PDFs of different individuals. The PDFs for one individual can be found in Supplementary Fig. 4. Our data reveals that individuals tend to vary the walking speed by changing step length while keeping the step frequency relatively constant (i.e., possibility two is true). This novel finding from extracting high fidelity trajectory data from a surveillance camera can contribute to the safety and comfort of the transportation systems. For example, transportation facilities, such as escalators and revolving gates, that may trigger step frequency change needs to be replaced by alternative designs that trigger step length change so that the walking can be safer and more comfortable.

## 3 Results and Discussion

We report a simple yet powerful crowd tracking method for extracting high-fidelity kinematic information (time-resolved position, velocity, acceleration, step frequency, and step length along the trajectory) that compares favorably with traditional methods. No information beyond four calibration points is required for analyzing the video data. In contrast to physical tracking methods that require camera calibration and a specific dress code for individuals in the crowd, our method works for almost all surveillance videos with reasonable spatiotemporal resolutions. With our technique, researchers can access scientific crowd data that is orders of magnitude larger than before. Relative to previous human tracking techniques focusing on resolving trajectory and velocity, our tracking technique also resolves higher-order statistics such as time-resolved acceleration, step frequency, and step length.

The simplicity of our technique is not based on compromised fidelity. To the contrary, we have shown that our method has high accuracy in coordinate transformation and calculating high order statistics. For instance, we show that our technique performs well for cameras with a spatial resolution of around 1 cm/pixel and a temporal resolution of around 30 Hz, resolutions similar to a large percentage of surveillance cameras. This technique opens the doors for extracting high fidelity kinematic information from the existing enormous amount of surveillance video data worldwide, serving to help meet pressing needs for understanding human mobility across fields, including but not limited to improving the safety and comfort of traffic systems, identifying precursors to coordinated acts of intentional physical violence in human crowds, and analyzing the behavior of crowds with social distance in a large gathering.

Our technique can help to improve the safety and comfort of the traffic system by better understanding the kinematics of pedestrians. We highlighted one possible application of the technique–how we can use surveillance camera videos to understand detailed single pedestrian statistics in crowds over a wide range of time spans and in a remote region, which is challenging, if not impossible, for traditional lab and field measurements. Our discovery of the strong correlation between an individual's speed change with step length exemplifies how time-resolved position, velocity, acceleration, step frequency, and step length can be exploited to reveal the hidden nuance of humans walking in different parts of transportation systems, such as a zebra crossing, which can offer the key knowledge for improving traffic systems for enhanced safety and comfort.

Moreover, detecting and understanding the kinematics of coordinated acts of intentional physical violence is an urgent need in many societies [24, 25]. It is currently not known how violent events emerged from human crowds that were originally seemingly peaceful. However, scientific measurement of such events is rare, if not missing, because one cannot predict the location and time of such an event. On the other hand, a volunteer-based mock study does not necessarily reflect the true scenario [45]. Lack of data imposes a major challenge for preventing coordinated acts of intentional physical violence in crowds. Nevertheless, surveillance cameras all over the world capture intentional physical violence accidentally, and there is a large volume of violent video data. With our technique, researchers can extract high fidelity crowd information from those videos and get the real kinematic information from actual coordinated acts of intentional physical violence. Our technique will lead to a vertical leap in physical violence study due to the ability to extract high fidelity kinematic data from surveillance videos.

In addition, intermittent and periodic outbreaks of infectious diseases have had profound and lasting effects on societies throughout human history, and social distancing is required during the pandemic, post-pandemic, and flu seasons. However, how individuals behave in socially distanced crowds in a large gathering is not understood to date [13, 17], without which we do not know how to facilitate the crowd in large events, such as parades and concerts. Our technique can help to reveal the hidden dynamics of socially distanced crowds based on the archived surveillance camera video during the pandemic.

Our method is ideally suited to service the aforementioned pressing needs for understanding human mobility across fields, as it enables high fidelity extraction of kinematic information of human trajectories straight from existing surveillance camera videos.

## Methods

### Direct Linear Transformation

DLT is an algorithm that solves a set of variables pertaining to cameras from a set of similarity relations. Traditional DLT is used for calibrating multiple cameras in three-dimensional space. With the assumption that all the humans are moving on a flat surface, we show that the 2D positions of the crowd can be determined by a single camera. This is a sound assumption in populated cities over a relatively small domain size. Our derivation shows that only four calibration points are required for calibrating cameras. The details are in Sec. Two-dimensional Direct Linear Transformation of the Supplementary Material.

### Longitudinal and transverse velocity and acceleration

Based on the aggregation of all trajectories, an averaged velocity vector $\mathbf{v}_{ave}$ can be calculated. Then, the longitudinal and transverse velocity/acceleration can be calculated by:

$$
\begin{aligned}
|\mathbf{b}| \cos \theta &= \frac{\mathbf{v}_{ave} \cdot \mathbf{b}}{|\mathbf{v}_{ave}||\mathbf{b}|} \\
|\mathbf{b}| \sin \theta &= \frac{|\mathbf{v}_{ave}|^2 |\mathbf{b}|^2 - (\mathbf{v}_{ave} \cdot \mathbf{b})^2}{|\mathbf{v}_{ave}|}
\end{aligned}
\quad (1)
$$

where $\mathbf{b}$ refers to an instantaneous velocity/acceleration vector and $\theta$ is the angle between $\mathbf{v}_{ave}$ and $\mathbf{b}$. So, the $\mathbf{b} \cos \theta$ is the longitudinal component while the $\mathbf{b} \sin \theta$ means the transverse magnitude.

### Auto-correlation of velocity and acceleration

The auto-correlation function shows the correlation between the velocity/acceleration of a particle at a reference time $t$ and some later time $t + \tau$ along the same trajectory. The definitions of velocity and acceleration auto-correlation function are written as:

$$R_{v_iv_i}(\tau) = \frac{\langle v_i(t)v_i(t+\tau)\rangle}{\langle v_i^2\rangle}$$
$$R_{a_ia_i}(\tau) = \frac{\langle a_i(t)a_i(t+\tau)\rangle}{\langle a_i^2\rangle} \qquad (2)$$

where the $\langle\cdot\rangle$ is the average over all the starting time $t$ with a fixed time lag $\tau$. $i$ is the index for different individuals.

## Gaussian Derivative Kernel

According to the convolution theorem $f * g' = f' * g$, where the $*$ is the convolution product, differentiating and filtering can be carried out spontaneously by applying a convolution product with the derivative of the Gaussian function to the objective signal. Gaussian function and its first and second derivative can be written as:

$$g(\tau) = \frac{1}{\sqrt{2\pi}\sigma}\exp\left(\frac{-\tau^2}{2\sigma^2}\right)$$
$$\frac{dg(\tau)}{d\tau} = \frac{-\tau}{\sqrt{2\pi}\sigma^3}\exp\left(\frac{-\tau^2}{2\sigma^2}\right) \qquad (3)$$
$$\frac{d^2g(\tau)}{d\tau^2} = \frac{1}{\sqrt{2\pi}\sigma^3}\left(\frac{\tau^2}{\sigma^2}-1\right)\exp\left(\frac{-\tau^2}{2\sigma^2}\right)$$

where $\sigma$ refers to the standard deviation. In practice, the Gaussian function needs to be discretized. Thus, a time window length, also called filter length, needs to be given. A proper filter length can filter out high-frequency noise while keeping the lower frequency signals. To find a proper filter length, the variation in characteristic velocity and acceleration with the length of filter length needs to be tested. For more details, please refers to Sec. Choosing the Cutoff Frequency for Trajectory Smoothing in the Supplementary Material.

## Acknowledgements

This research was partially supported by Pitt Momentum Funds at the University of Pittsburgh. We would like to thank Wenhe Ma and Hao Yu for helping with code execution and experiments.

## Author contributions

L.F. conceived the method and devised the research. L.F. performed the experiments. Z.L. and L.F. performed data analysis and wrote the manuscripts.

## Additional information

All the measurements contain no ethical issues, as confirmed by the institutional review board at the University of Pittsburgh.